\documentclass{kluwer}    

\usepackage{graphicx}

\newdisplay{guess}{Conjecture}

\begin{document}
\begin{article}

\begin{opening}         

\title{Star Formation in a Multi-Phase ISM} 

\author{Stefan \surname{Harfst}}
\author{Christian \surname{Theis}}
\author{Gerhard \surname{Hensler}}
\institute{Institute for Theoretical Physics and Astrophysics, University Kiel, Germany}

\runningauthor{S. Harfst et al.}
\runningtitle{Star Formation in a Multi-Phase ISM}

\begin{abstract}
We present a 3d code for the dynamical evolution of a multi-phase
interstellar medium (ISM) coupled to stars via star formation (SF) and
feedback processes. The multi-phase ISM consists of clouds (sticky
particles) and diffuse gas (SPH): exchange of matter, energy and
momentum is achieved by drag (due to ram pressure) and condensation or
evaporation processes. The cycle of matter is completed by SF and
feedback by SNe and PNe. A SF scheme based on a variable SF efficiency
as proposed by \inlinecite{EE97} is presented. For a Milky Way type
galaxy we get a SF rate of $\sim$1\,${\rm M}_\odot\,{\rm yr}^{-1}$
with an average SF efficiency of $\sim$5\%.

\end{abstract}


\end{opening}           

\section{Introduction}  
\label{harfst_sec_intro}

So far in 3d models of galaxies the ISM is mostly described either as
a diffuse phase by SPH (e.g.\ \opencite{HK89}) or as a clumpy phase by
sticky particles (e.g.\ \opencite{TH93}). Alternatively,
chemo-dynamical models use a multi-phase ISM, but were usually
restricted due to spherical or axisymmetric systems (e.g.\
\opencite{TBH92}; \opencite{SHT97}).  In order to extend the
chemo-dynamical models to three dimensions we combine both treatments
in a particle based code: the hot diffuse gas phase is described by a
SPH formalism, whereas the cold molecular clouds are represented by
sticky particles. The coupling between the gaseous phases is achieved
by condensation/evaporation (C/E) and by a drag force due to ram
pressure. Energy is dissipated by cloud-cloud-collisions or by
radiative cooling. Furthermore, stars are formed in clouds and the
stars return mass and energy to the gas by feedback processes
(supernovae, planetary nebulae).

In SPH codes the SF is generally based on the Schmidt law, i.e.\ the
SF rate depends on gas density and a characteristic time scale. For
sticky particles the SF can be coupled to cloud-cloud collisions, or a
single cloud forms stars with a constant SF efficiency. Here, a SF
scheme using a different approach is presented: the process of SF is
treated individually for each cloud. The SF efficiency is dependent on
local properties of the ISM and the star forming clouds, thereby
enabling self-regulation of SF by feedback.



\section{Numerical Treatment}
\label{harfst_sec_code}

The gravitational force for all particles is calculated with a TREE
algorithm. A dark matter halo can be modeled as an external, static
potential (static halo) or by means of particles (life halo). 
The integration of the system
is done with a leap-frog scheme.  Individual time steps are used for
each particle.

The drag force and the mass exchange rates for C/E are calculated
following \inlinecite{CMO81} using a mass-radius-relation for the
clouds based on observations (e.g.\ \opencite{RS88}). Effects of both
processes are individually determined for each cloud using the local
density, temperature and velocity of the hot gas.  Energy can be
dissipated by inelastic cloud-cloud collisions \cite{TH93} and by
radiative cooling
\cite{BH89}.

\begin{figure}[t]
\centerline{\includegraphics[width=\textwidth]{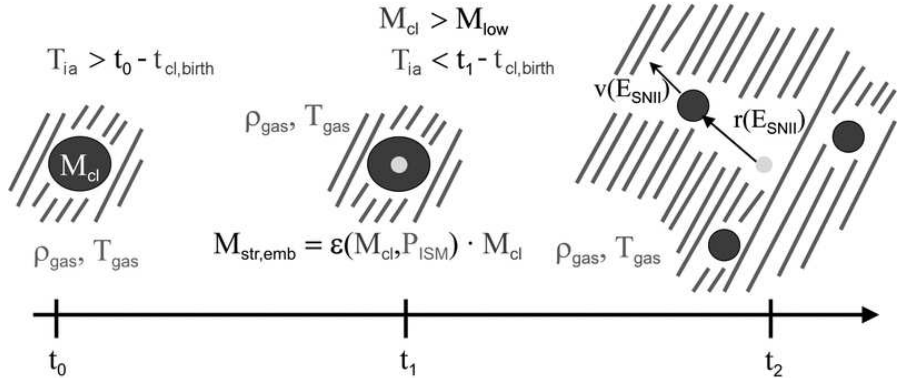}}
\caption{
The SF scheme (see text for details)}
\label{harfst_fig1}
\end{figure}

Clouds can transform some of their mass into new stars. In
Fig. \ref{harfst_fig1} this process is drawn schematically. Each cloud
is inactive for a given time ${\rm T}_{\rm ia}$ (${\rm t}_0$). An
embedded star cluster is then formed (${\rm t}_1$) with the SF
efficiency $\epsilon$ being a function of cloud mass and gas pressure
according to \inlinecite{EE97}. The energy released by SNe is
calculated using a multi-power law IMF \cite{KTG93}. This energy
injection disrupts the cloud into smaller fragments (${\rm t}_2$). The
time for disruption ${\rm T}_{\rm dis}$ (${\rm t}_2 - {\rm t}_1$ in
Fig. \ref{harfst_fig1}) is determined from the energy input: a
self-similar solution is used to calculate the SN shell expansion. It
is assumed that the cloud disrupts when the shell radius equals the
cloud radius. The fragments are then placed symmetrically on the shell
with velocities corresponding to the expansion velocity at ${\rm
T}_{\rm dis}$. The energy not used for cloud disruption and the mass
ejected by SNe is returned to the hot gas phase (SPH
particles). Additionally, the mass returned by PNe is added to
surrounding cloud particles.


\section{The model and results}
\label{harfst_sec_res}

Initially a galaxy similar to the Milky Way was set up using the
galaxy building package described by \inlinecite{KD95}. A cloudy
gas phase is added by treating every tenth particle from the stellar
disk as a cloud. The total mass in clouds is ${\rm M}_{\rm cl,tot}
\approx 3.4\cdot10^9\,{\rm M}_{\odot}$ and each cloud is assigned a
time of inactivity ${\rm T}_{\rm ia}$ between 0 and 200\,Myr. Finally,
a hot gas halo with a total mass of ${\rm M}_{\rm hot,tot} \approx
2\cdot10^8\,{\rm M}_{\odot}$ is added, which is designed to be
in hydrostatic equilibrium in the halo potential.

\begin{figure}[!t]
\tabcapfont
\centerline{
\begin{tabular}{c@{\hspace{6pt}}c}
\includegraphics[width=170pt]{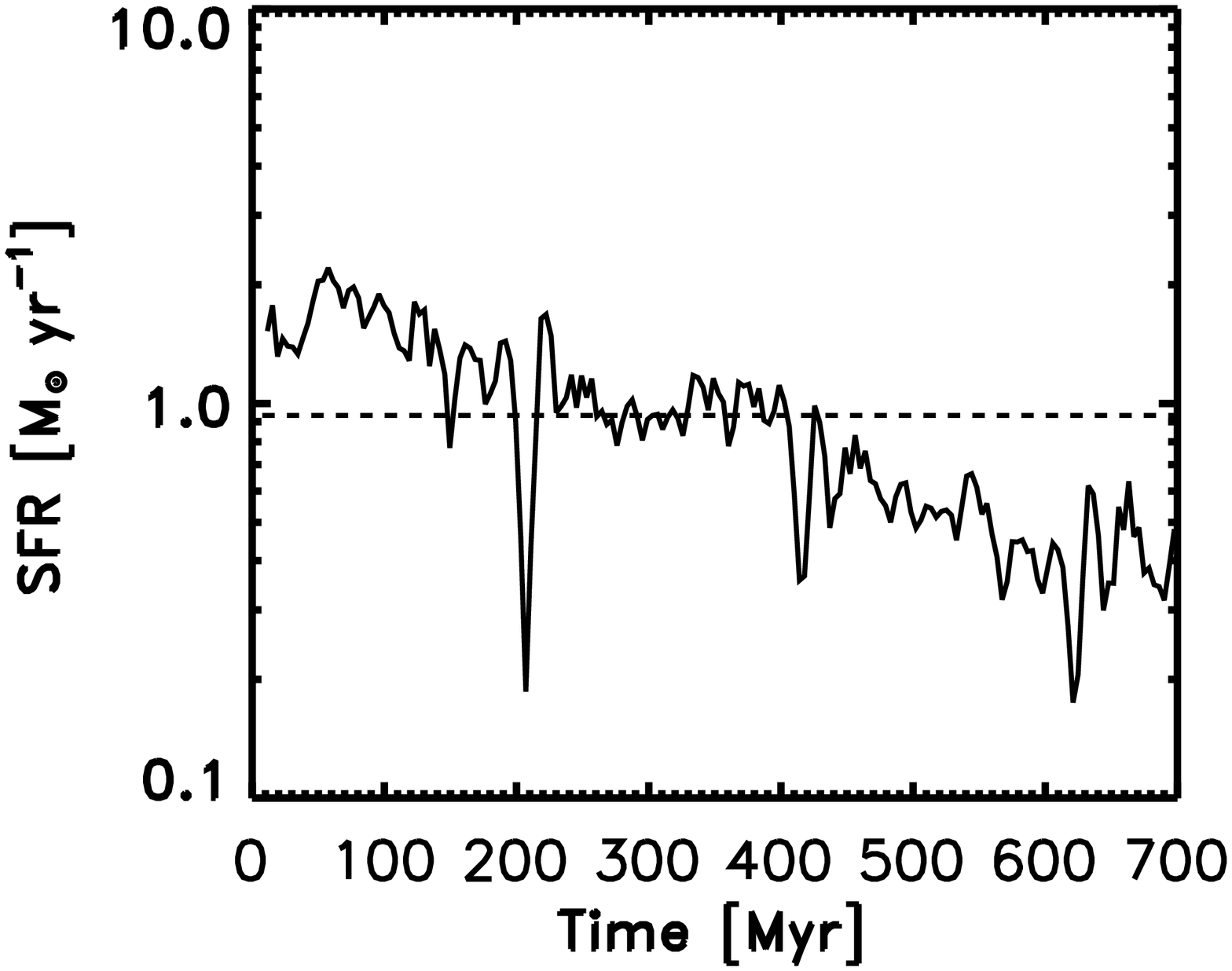} &
\includegraphics[width=170pt]{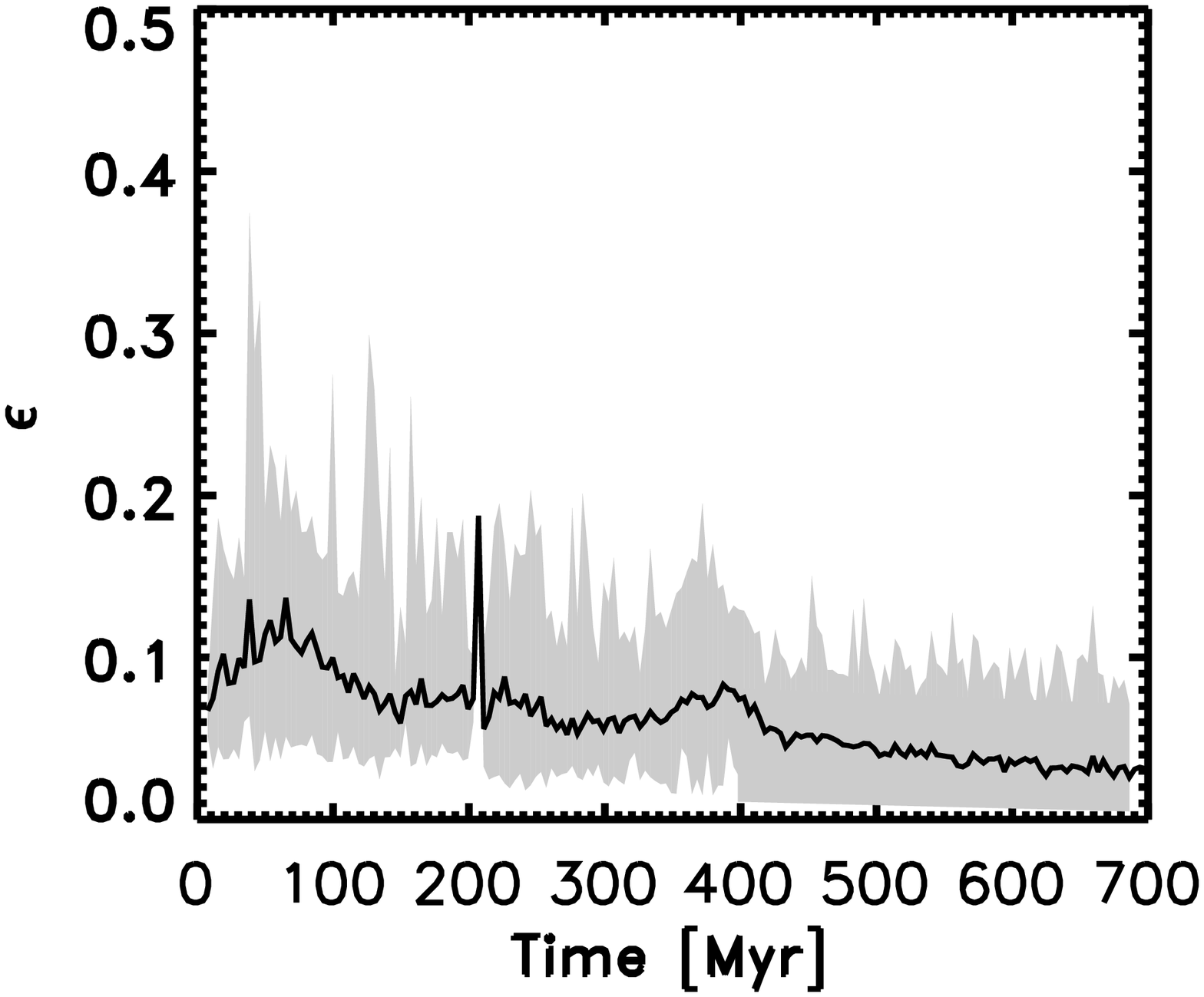} \\
a)~~ Star formation rate & b)~~ Star formation efficiency
\end{tabular}}
\caption{
In the left plot the temporal evolution (solid line) and the average
(dashed line) of the SF rate are shown for a Milky Way type
galaxy. The right plot shows the average SF efficiency (solid line)
and the range of individual SF efficiencies (shaded area) with time
for the same model.
}
\label{harfst_fig2}
\end{figure}

In Fig. \ref{harfst_fig2}a the resulting SF rate is shown, which is on
average $\sim$1\,${\rm M}_{\odot}\,{\rm yr}^{-1}$. The SF rate is slowly
decreasing with time due to 1) the decreasing mass in the cloudy phase
of the ISM and 2) the decreasing average SF efficiency
(Fig. \ref{harfst_fig2}b). The decrease of $\epsilon$ is correlated
with a decreasing ${\rm M}_{\rm cl}$ as expected from
\inlinecite{EE97}. After 700\,Myr the average SF efficiency is less
than 5\% and the maximum efficiency is always lower than 20\%. From
such a low SF efficiency we do not expect any non-induced globular
cluster formation in the disk.

\begin{figure}[!t]
\tabcapfont
\centerline{
\begin{tabular}{c@{\hspace{6pt}}c}
\includegraphics[width=170pt]{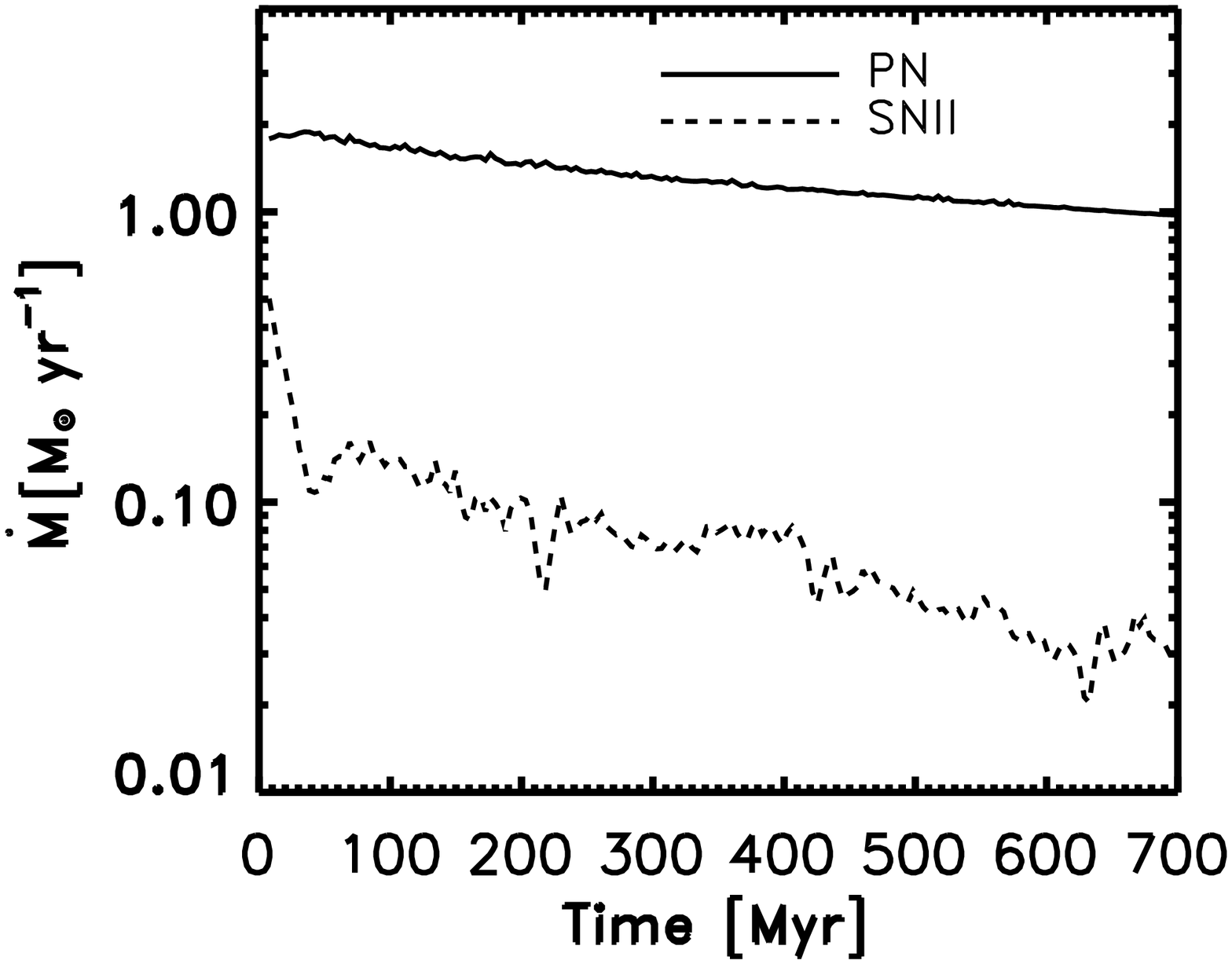} &
\includegraphics[width=170pt]{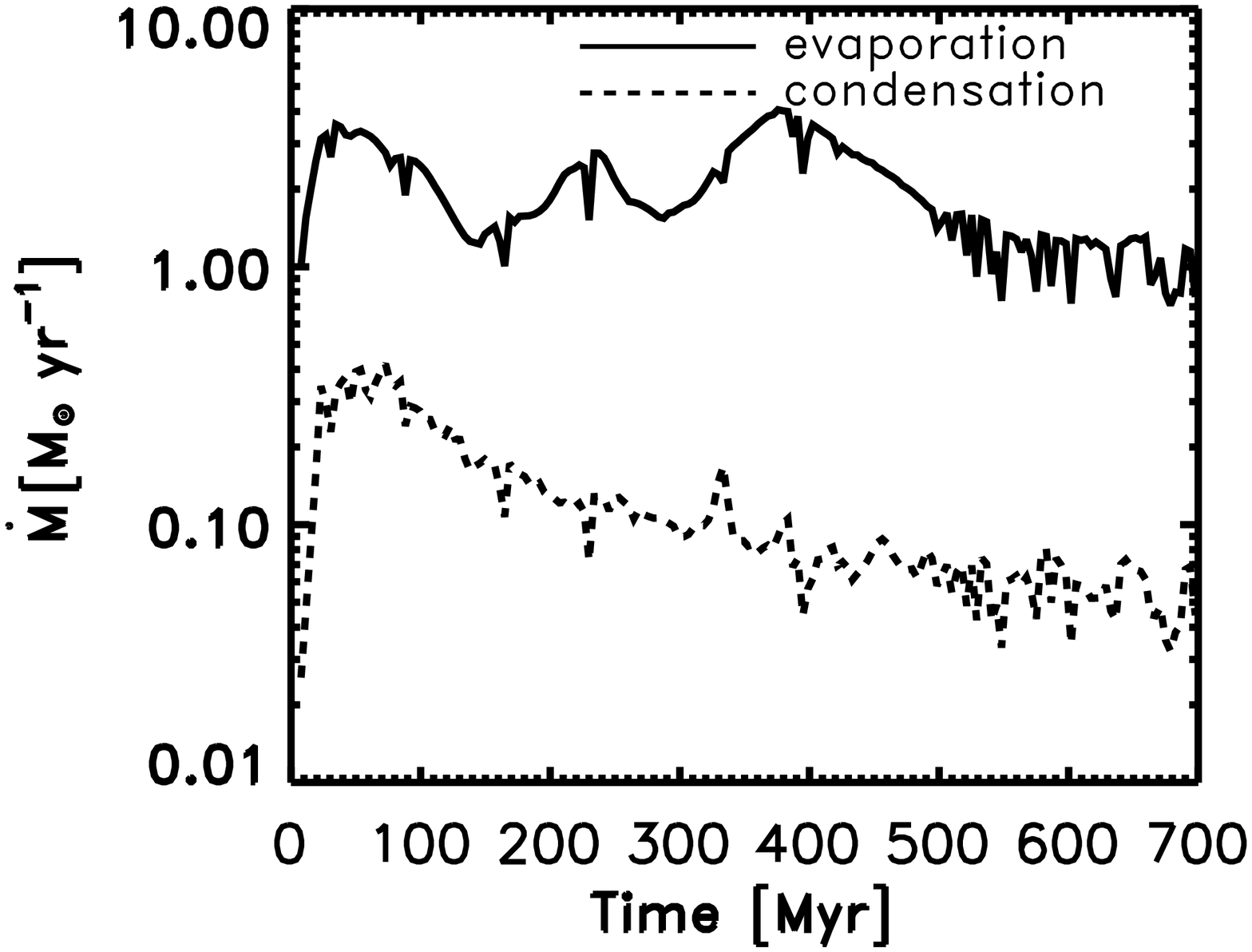} \\
a)~~ Feedback & b)~~ Condensation and evaporation
\end{tabular}}
\caption{
Mass flow rates. The left plot shows the temporal evolution of the
mass return rate by PNe (solid) and SNe (dashed). In the right plot
condensation (dashed) and evaporation (solid) rates are shown.
}
\label{harfst_fig3}
\end{figure}

In Fig. \ref{harfst_fig3}a the mass return rate due to stellar death
is shown: the SNe rate is about 10\% of the SF rate and in agreement
with the mass fraction of massive stars. The mass return rate by PNe
is approximately equal to the SF rate due to a contribution of older
stars in the initial model.

Condensation and evaporation rates are shown in Fig
\ref{harfst_fig3}b. The evaporation rate is about 10 times higher than
the condensation rate because the temperature of the hot gas amounts
to several $10^6\,$K. While the SF establishes an equilibrium on a
rather short time scale C/E have not reached an equilibrium state
after 700\,Myr.  This is in agreement with the analytic work of
\inlinecite{KTH98}. They predict an equilibrium state only after a few
condensation time scales of evolution, which is a few Gyr in this model.


\acknowledgements
This work is supported by the {\it Deutsche Forsch\-ungs\-gemein\-schaft (DFG)} 
under the grant TH-511/2-3.

\bibliographystyle{klunamed}           
\bibliography{harfst}

\end{article}
\end{document}